\documentstyle[numreferences]{kluwer}    

\let\oldverbatim\verbatim
\renewcommand{\verbatim}{\expandafter\small\oldverbatim}
\runningtitle{Workshop On Gravitational Waves}
\runningauthor{B.R.Iyer and K.D.Kokkotas}
\begin{opening}
\title{ Workshop on Gravitational Waves
\footnote{Report of the {\em Workshop on Gravitational Waves} 
held during the ICGC-95 in Puna,  12--19 December 1995. }
}
\author{Bala R. \surname{Iyer}}
\institute{Raman Research Institute  \\ 
           Bangalore 560 080 \\
           INDIA}
\author{Kostas \surname{D. Kokkotas}\thanks{Permanent address:
Department of Physics, Aristotle University of Thessaloniki,
Thessaloniki 54006,GREECE}}
\institute{Max-Planck-Society, Research-Unit Theory of Gravitation\\
University of Jena, Jena D-07743\\
GERMANY}

\end{opening}

\begin{ao}
Prof. B.R.Iyer\\
Raman Research Institute\\
Bangalore 560 080\\
INDIA
\end{ao}                   

\begin{document}
  



\section{Introduction} 
According to general relativity any change in the gravitational 
field of a compact gravitating system like neutron star or black hole,
will produce a ripple on the spacetime which will propagate with 
the speed of light. This is a {\em gravitational wave}. 
Although gravitational waves have been predicted from the very 
early days of general relativity they have not yet been detected 
directly. 
Nevertheless, their indirect influence on the 
evolution of compact binary systems has been observed almost
20 years ago, and the observational results agree to remarkable 
accuracy with the theoretical predictions.
 
Almost thirty five years ago attempts were made to detect these tiny ripples 
in the spacetime by constructing 
gravitational wave detectors in the form of resonant bars.
Today these detectors are either narrowband resonant bars or spheres
or broadband laser interferometers.
The resonant detectors are already in operation and they 
have achieved sensitivities close to their limits, while the laser 
interferometers (LIGO, VIRGO, GEO600) are in the construction 
phase and  expected  to become operational before the end of 
this century. 
The next generation of gravitational wave detectors includes space 
experiments  like LISA accepted by ESA as a possible cornerstone
mission. 
 
The direct detection of the gravitational waves will not  only be a great 
triumph for Einstein's theory of relativity but  is expected 
to signal a new era in astronomy by opening a non electromagnetic 
window of observation. 
By a network of at least three laser interferometers one will be 
able to determine the distance of a coalescing binary system, 
its position in the sky, the masses of the members of the system
and orbital frequency at final merger of the system. 
This information in turn will enable the determination of the Hubble constant
\cite{BFS86}, distribution of masses of neutron stars and black holes
and constraints on the  neutron star equation of state. 

The era of {\em gravitational wave astronomy} is hopefully not too much
in the far future and there is an international  collaboration between 
experts from various different backgrounds to solve the relevant
problems and march to  this new epoch. 
We see  combined work from various groups around 
the world in the fields of  numerical relativity, 
techniques in  data analysis,  approximation 
methods 
 and importantly
 an exchange of ideas and methods between
technical groups  around the various detectors both bar and 
interferometer. 
 
In the workshop for gravitational waves there were talks related to 
the theory of the gravitational waves, the  methods to be used to analyse the 
data and  extract 
information  as well as 
presentations on  the status of  experimental work on 
bar and laser interferometric detectors. 
 The most promising source of gravitational waves is the binary 
neutron star or black hole coalescence. Though most of the 
talks   
were focused on these systems, nevertheless, stellar collapse 
to neutron stars or black holes as also  
gravitational waves from pulsars were
also discussed.

The workshop was divided into two sessions: A discussion
session   to summarise present status and future
directions in some areas we considered important;
the other  a session  of contributed papers  to present interesting
work in progress.
The list of discussion topics is not exhaustive but indicative. 
The endeavour to detect gravitational waves probably pushes to the
extreme every aspect that it involves. Technology of all kinds, vacuum,
optics, isolation systems, noise control.
It also has need to push theory to
orders of accuracy not available today. For once in general relativity
experiment drives the theory!  We begin with theoretical aspects and look
at the inspiralling binary system where the frontier lies at third- 
post-Newtonian (3PN) order. 
One needs to address the issue of generation on one hand and equation of
motion on the other. The best calculation in the restricted two body
limit is at 4PN. What next? To go beyond the approximation and
perturbation techniques to the fully general relativistic regime as of
date there is only the numerical route.
The ``grand challenge'' is the problem
of  black hole collisions in three dimensions but the testing grounds for
conceptual and numerical issues are one dimensional and two dimensional
problems. All this work is driven by  the need to have high phasing
accuracy in the gravitational wave templates. Data analysis aspects need
to be carefully assessed to separate what we really know from what
we think we know, the folklore from the facts. Though the logo
of our community could well be the inspiralling and coalescing
binary it is important to explore weaker sources like pulsars.
And finally even at this late hour reflect whether we
``interfere'' or ``resonate''.

\section{Inspiralling Compact Binaries : Problems in the 3PN Program}
One of the most urgent challenges on the theoretical front 
in  the PN programme is the 3PN generation and this was discussed 
by {\sf Luc Blanchet}.
A challenging problem was to predict the time evolution of the orbital
phase of an inspiralling compact binary, as due to the reaction to the 
emission of gravitational waves. Such prediction was needed for the
future observations of the LIGO and VIRGO detectors. The demanded
precision should be extremely high, namely it should take into 
account many high-order relativistic or PN corrections 
\cite{3mn,CFPS94,P95}.

The computation generally relied on an energy balance argument which
equated the decrease of binding energy of the binary and (minus) the 
energy flux generated by the binary in the form of gravitational waves. 
The validity of this argument had been proved up to now only to 1.5 
post-Newtonian (1.5PN) order, or order $c^{-3}$ when the speed of light 
$c \to \infty$ \cite{IW93,IW95,B93,B95reac,BD88}. 
A general and important 
problem was to prove that this argument stayed correct to much higher order 
than 1.5PN. 

The binding energy of the binary resulted from the equations of motion
of the binary which were known presently up to 2.5PN order 
\cite{DD81a,D83a}. 
On the other hand the energy flux at infinity was also known at 2.5PN order 
\cite{B95,BDI95,B96pn,WW95}. 
{\it Assuming} that the energy balance argument 
was valid at this order, the orbital phase was therefore known presently
up to 2.5PN.

To 3PN order the following problems must be (or have already been) solved.
\begin{itemize}
\item The multipole moments of the binary system are needed at 3PN. Some 
general expressions for the moments as integrals over an arbitrary source 
(in principle), and formally valid up to any post-Newtonian orders, 
are known. However the proof of the validity of these expressions 
still needed some clarification. The reduction of these expressions to 
binary systems was a hard computational task (which should be done partly 
on the computer). Many intricate integrals are to be evaluated. It
was not clear that all of them could be obtained analytically.
\item The nonlinear effects in the wave zone gave an important observational
contribution. At 3PN order in the energy flux nonlinear terms were known 
to be composed solely of ``tails of tails", which are the tails of waves 
generated by curvature scattering of the (quadratically nonlinear) tails 
themselves. This effect was cubically nonlinear. It was presently under 
control.
\item The equations of motion of the binary should be generalized from 2.5PN
to 3PN order, at least for circular orbits. The equations of motion were
needed both for computing the binding energy and for reducing the time 
derivatives when computing the energy flux. This problem was a difficult 
one in part because one needed to justify at this order the use of delta 
functions to represent the compact objects.
\end{itemize}
These problems when solved should give a result in perfect agreement, 
in the test mass limit for one body, with the result of black-hole 
perturbation theory \cite{Sasa94,TSasa94}.

This was followed by a remark  by  {\sf  Piotr Jaranowski} on
the technical hurdles in the computation of the 3PN equation of motion 
which
was  in progress in collaboration with Gerhard Sch\"afer. 
They  were currently calculating 3PN Hamiltonian for the two
body point-particle system in the framework of ADM formalism.

They had encountered some serious ambiguities due to highly singular structure
of the integrals involved in the calculation. There were terms for which
Hadamard's ``partie finie" procedure gave the result different from that
obtained by using the Riesz's kernel representation of the Dirac delta
distribution.

They had also recognized that regularizing integrals by means of multiplying an
integrand by $r_1^\alpha r_2^\beta$ was of restricted validity, because it did
not give zero for some integrals being full divergences. They should be zero
due to the obvious identity 
$\left( \nabla_1 + \nabla_2 \right) f( {\bf r}_{12} ) = 0$.

\section{  Black Hole Perturbation  Approach}
{\sf Misao Sasaki} reviewed the recent developments in the black-hole 
perturbation approach to
gravitational waves from coalescing binaries.
This approach, though valid only in the limit when one of the two bodies is 
much heavier than the other in the strict sense, has been shown to be very 
useful in analyzing higher post-Newtonian features of the gravitational
radiation and the orbital evolution of a binary in a complementary
manner as compared to the standard post-Newtonian approach. 

The highest PN order analytically achieved so far is 4PN for
a non-spinning particle \cite{jap1} and 2.5PN for a spinning particle 
\cite{jap2}, 
both in the case of circular orbits around a Kerr black hole.
Although the calculations would become increasingly complicated, there
seemed no fundamental difficulty to go to still higher PN orders  for
circular orbits or those allowing perturbative expansion from a circular
orbit. A matter of discussion would be whether it would be meaningful or useful
to do so. In this respect, one interesting issue was the convergence
property of PN expansion. This can be examined to good accuracy by
comparing the results of analytical expansions with those obtained
numerically which are `exact' up to numerical precision.

A more meaningful and perhaps interesting problem was the radiation
reaction, since one would be able to gain useful insights into the
radiation reaction mechanism in relativistic situations.
He mentioned some recent attempts to derive the radiation
reaction force term in certain restricted situations. But the status of
the problem seemed quite obscure at the moment in his opinion. The main
reason was that we don't have a well-justified method to regularize the
divergence associated with the point-mass.

Finally, he mentioned the issues of post-Teukolsky expansion and the
perturbation approach for background spacetimes other than a black hole.
Although there were no good ideas, he emphasized that if we found a way to
handle these issues, the perturbation approach would become a much more 
powerful tool.

{\sf Jorge Pullin} spoke about the ``Close approximation" techniques for dealing with
the collision of two black holes. The idea was that if one considered
collisions where the two black holes started close to each other
one could approximate the situation by a single distorted black hole
and apply black hole perturbation theory to the problem. 
This was shown to work very well in the problem of the head-on
collision of two black holes \cite{pull1}. The challenge next was to apply it
to other collision scenarios. For that purpose two things were needed
a) initial data and b) some estimate of the degree of trustworthiness
of the results. 

In order to attack a) they proposed to solve the initial value problem in
the close approximation. In this approximation one could find simple yet
accurate solutions of the Hamiltonian and momentum constraint for
situations like black holes with spins and linear momenta.
At the moment they \cite{pull3} were studying
the case of collisions with momentum and the counterrotating collision
of black holes with spin \cite{pull4}.  This latter case could be compared with
the full numerical simulations of Brandt and Seidel \cite{pull5}. The case of
in-spiralling collision with small total angular momentum could also be
treated as a perturbation of Schwarzschild and this was their next
step. One could refine this by studying perturbations of Kerr using the
Teukolsky equation but this was more involved, although it was
feasible. Others are initial data of different symmetries \cite{pull6}.

On the issue of how trustworthy the formalism was, one could construct
certain empirical criteria, like how much did the linearized initial
data violate the exact Hamiltonian constraint \cite{pull7}. These criteria were
useful as quick tests but may not be accurate enough.

\section{Numerical Relativity Challenges}

{\sf Edward Seidel} discussed
 briefly a number of important challenges and problems for
numerical relativity to solve in the coming decade.  He felt that although
significant progress had been made over the last decade, general, 3D,
long term evolutions of systems governed by the Einstein equations were
still very difficult.\\
Numerical relativity has been an active
field of research for about 30 years.  As analytic techniques were
still unable to penetrate the full solution space of Einstein's
equations for nonlinear, time dependent systems in 3D, numerical
solutions were presently our best hope for unlocking their predictive
power for systems without symmetries.  Significant progress had been
made in the last decade in numerical relativity, leading to evolutions
of distorted and colliding black holes in 2D and 3D\cite{Seidel96a},
as well as pure gravitational wave spacetimes, and self-gravitating
scalar fields.  The entirely new field of critical phenomena came from
numerical relativity.  It was a very active and exciting area of
research, but there was still much work to be done!

In many areas there was a pressing need for both theoretical and
numerical study.  Particularly important was the interplay between
theoretical ideas and numerical implementation.  Although much of the
current effort in numerical relativity was directed towards 3D, many of
the numerical studies could be carried out in simpler spacetimes in 1D
and 2D, which were now accessible on powerful workstations.  Techniques
developed on these more accessible problems could then in many cases be
applied to the more complicated 3D cases.\\
A very basic problem faced by all numerical relativists was the long
term, stable evolution of the Einstein equations.  At present, all
black hole evolutions in 2D and 3D were plagued by numerical
instabilities at late times ($t\approx 200M$ or less, where $M$ was the
mass of the black hole)\cite{Seidel96b,Ibanez96a}.  But even in very
weakly gravitating systems, such as low amplitude gravitational waves,
numerical instabilities could develop after a fairly short
time\cite{Anninos96b}.
Some of these instabilities are
related to finite differencing across large gradients that develop in
metric functions near black holes, some of them are related to the
drifting of coordinate lines in 3D, and some of them are unknown in
origin.

There were many lines of research that were needed to address these
problems.
For example, which lapse and shift are needed in 3D was
an open question. Good choices would likely be geometric in nature
(e.g., minimal distortion shift\cite{York79}), would minimize
coordinate drift and also the stretching and shearing of metric
functions that could lead to problems.
The freedom in the lapse has
traditionally been used to avoid developing spacetime
singularities\cite{Anninos94e}, but it can also be used to control
certain properties of the metric.  Good choices of slicing and shift
conditions in 3D numerical relativity need to be explored.
Another important attack against instabilities that develop in
numerical evolution was the finite differencing of the equations
themselves.  In the standard 3+1, ADM approach to numerical
relativity\cite{York79}, the equations were extremely complicated,
but in hydrodynamics the equations took on a special form where
derivatives appear in the so-called ``flux'' terms.
The physical interpretation of
these terms lead to special finite difference operators that preserved
important physical characteristics of the solution.  In cases where
the equations are hyperbolic, the system could be diagonalized and the
eigenfields themselves could be evolved.
Recently,
the Einstein equations had been cast in this first order, flux
conservative hyperbolic form (FOFCH)\cite{Bona92,Bona94b}.  Dramatic
results had been obtained in spherical symmetry, and 3D codes based
on this hyperbolic formalism, and others\cite{Abrahams95a}, were under
construction.  But this line of research was just beginning, and much
work remained to be done to understand the best way to take advantage
of these hyperbolic formulations.\\
Black holes were among the hardest systems to
handle in numerical relativity, even in the stationary case where the
solutions were known analytically!  Axisymmetric calculations had been
carried out for the last decade, but at present, we were still limited
to rather crude evolutions in 3D, even in cases with a high degree of
symmetry\cite{Seidel96a,Seidel96b}.  For the more general case of two
spiralling black holes, we were far from the final solution, and much
work was needed to be done to achieve that goal.

There was no room to detail all the many areas of research related to
black hole evolutions, but he would simply mention the major issues
and give references to current work.  Recent work on using ``apparent
horizon boundary conditions''\cite{Seidel92a,Anninos94e,Scheel94}
showed that it may be possible to handle singularities in numerical
calculations by removing the region inside a horizon from the
calculation.  This technique allowed one to avoid using pathological
singularity avoiding time slicings, and had worked well in spherical
symmetry.  However, a stable, long term evolution for a true 3D black
hole had not yet been tried, and much remained to be done to perfect
the technique. In 3D spherical black holes
\cite{Anninos94c},
and some axisymmetric systems had been studied, but no spin or
angular orbital momentum had been attempted.  He expected that this area
of research would keep people busy for quite some time to come!\\
General relativistic hydrodynamics would be a very exciting area, as it
encompassed most of astrophysics.  Many important problems in
astrophysics, such as accretion onto black holes, mergers of neutron
stars with other neutron stars or with black holes, stellar core
collapse, supernovae, etc., could only be fully treated through a fully
self consistent approach to solving the coupled Einstein and
hydrodynamic equations.  However, very little had been done in this
area until now, primarily because of the difficulties of solving the
(left hand side of) the Einstein equations\cite{Ibanez96a}.  
This was a rich area of relativity and
astrophysics and would get much more attention in the coming decade.\\
Pure gravitational waves provided a laboratory for studying the full
nonlinear theory without complications associated with matter fields
or singularities in the initial data.  This area of research was for
the most part uncharted territory.  The general behavior of 3D strong
gravitational waves, including for example gravitational geons and the
formation of singularities, was unknown.
Many questions about waves in more
general 3D spacetimes were waiting to be studied.\\
The interplay of analytic studies with numerical relativity would
become even more important as numerical simulations became more
sophisticated.  We would need a full set of tools by which to analyze and
understand the results of the simulations.  An excellent example of
this was the recent work using using perturbation theory to explore the
dynamics of black hole collisions (See e.g., \cite{Price94b} and
references therein.)  Also, numerical and theoretical studies of
apparent and event horizons had recently led to a better
understanding of their dynamics (See, e.g., \cite{Seidel96b} and
references therein.) There was much more to be done along these lines,
and others, such as the use of traditionally analytic tools,
including Riemann Invariants, principal null directions, etc.  This
should be an area where more analytically inclined researchers could
make an essential contribution to numerical relativity.\\
In conclusion, numerical relativity had made great strides
during the last decade, but many challenges lay ahead.  There was room
for groups from all areas to contribute, in both theoretical and
numerical areas.  For those interested in getting started in numerical
relativity, 1D, 2D, and even 3D sample codes were available from the
NCSA relativity group WWW server at http://jean-luc.ncsa.uiuc.edu.

\section{Data Analysis: What do we agree upon?}

In his presentation {\sf B.S. Sathyaprakash} presented a critical
resume of the data analysis endeavours.
He pointed out  aspects of gravitational wave data
analysis that had in his view,
gained consensus amongst people working in those areas and indicated problems
that needed further exploration. The sources discussed include coalescing
compact binaries, non-spherical neutron stars, stochastic background 
and burst sources.

The first in our list of candidate sources was the coalescing compact
binary. It was now reasonably well agreed upon that matched filtering
would be employed to extract chirp signals buried in noisy data. The
number of templates required to span the parameter space of the inspiral
wave form sensitively depended on the lower limit in masses of
binaries \cite {sd91} that we intended to search as well as the
post-Newtonian order of the template of wave forms\cite {bjo96,taa96,bss96}.
By now it was well established that
the restricted post-Newtonian wave forms, only incorporating the
phase corrections to the wave form, were good enough for detection\cite {3mn}.

Amplitude corrections at worst induced biases in the estimation of
parameters and these could be taken care in off-line analysis.
It was however not known how important were the eccentricity induced
modulations of the wave form.
For a search covering the range $[0.5,20]M_\odot$ of the total
mass of the binary one would need as many as $2 \times 10^4$ post-Newtonian
templates and a computing power of 2 GFLOPS for an on-line search while
the corresponding numbers for $[1.0,20]M_\odot$ were $3500$ post-Newtonian
templates and 250 MFLOPS computer power\cite {bjo96,bss96}.  
A parameter called the chirp time,
which was a certain combination of the masses of the two stars had been
found to be a good parameter for the purpose of making a choice of templates
\cite {bss94}.
It remained to be seen as to what was a good second parameter. 

There had been several attempts to find a solution to the inverse problem of
reconstructing the full wave form (the two independent polarisation amplitudes,
direction of the incoming wave form and the parameters of the wave form)
with a knowledge of the response function from a network of three or more
detectors\cite {dt88,GT89,JK94,JKKT96},
but an optimal solution was still lacking. In the area of
parameter estimation Cramer-Rao error bounds were now available 
at various post-Newtonian orders\cite {FC93,CF94,PW95,KKS95,bsd95,bsd96}.
However, we needed estimates of
biases induced in estimation when inaccurate templates were employed
in detection \cite{bss96}.
Moreover, Cramer-Rao bounds were not useful since
they did not tightly constrain the errors. Recent work employing Monte
Carlo simulations had shown that errors in the measurement of parameters
were typically  larger by a factor of 2--3 than Cramer-Rao bounds
\cite{bsd95,bsd96,kkt94}.
There was considerable work on testing theories and models which had shown
that gravitation theories \cite{bs95}
and cosmological models \cite {lsf96} could be constrained at high confidence
levels with observation of few to many events.
Further work was needed in finding how well astrophysical models could be
tested.

Second in our list of sources were the sources that emitted periodic 
gravitational waves continuosly, namely
spinning, non-spherical neutron stars that could or could not
be observable electromagnetically. The extent of amplitude and frequency
modulation caused due to Earth's motion relative to the solar system
barycentre had been worked out\cite {kj95,gsj95} and attempts
had been made to compute the Fourier transform of the modulated wave
form \cite{kj95}.
The signal was
envisaged to be detected by Doppler de-modulating and Fourier transforming
data that was typically several months long \cite{bfs91}.
The problem was that de-modulation depended on the direction to the source.
Present estimates of the number of patches in the sky that needed to be
corrected for modulation effects (so that there was no appreciable loss in
signal-to-noise ratio for all sources located in a given patch) was in excess
of $10^{13}$ while searching a year's worth of data\cite {gsj95}.
Computing power needed to carry out
an on-line, all-sky, all-frequency search was certainly out of reach
even by the standard of computers that were expected to become available
towards the turn of the century.
It would however be possible to do
on-line search for known pulsars and possibly all-sky, all-frequency
search for pulsars in a week's worth of data.
However, large proper motions observed in the case of some pulsars
caused additional drop in signal-to-noise ratio and it needed to be
explored as to what was the magnitude of this motion
and to what extent this could be corrected for.
There had not been much work on the estimation of parameters such as the
extent of nonsphericity, neutron star magnetic fields, direction to the
source, etc.

Stochastic background of gravitational waves was the next in the list.
Here detection was envisaged by cross correlating data from two nearby
interferometric detectors or an interferometric detector in a
narrow band operation and a cryogenic bar with a compatible
operating frequency and sensitivity \cite{bfs91}.
In either case the signal-to-noise
ratio was enhanced in proportion to $T^{1/4}$ where $T$ was the observing
time. There had been estimates of the amplitude and power spectrum
of background gravitational waves produced in models of inflation but
how such models may be constrained with the aid of data from gravitational
wave detectors remained to be estimated. We did not know how the noise
intrinsic to the detector contaminated the cosmological background. Monte
Carlo simulations of detection of the background would greatly help in
learning more about these sources.

The last in the list of sources was the burst of radiation emitted in
a supernova event and other short lived burst sources. It could be
possible to dig out bursts buried in noise by looking for generic
features such as a characteristic frequency and a ring down time.
Unfortunately there was a great variety of wave forms predicted in
a supernova depending on the nature and extent of asymmetry in the
collapse, equation of state of gravitating matter, details of nuclear
astrophysical processes, etc. and thus supernova events fell into the
category of unknown sources as far as their detection was considered.
It could be that they could only be detected by cross correlating data
from different nearby detectors or looking at high-sigma events in the
time series. Much work was needed to be done with regard to data analysis of
bursts.

\section { On the Detection of Gravitational Waves from Pulsars}

{\sf Sanjeev Dhurandhar} summarised the status of the studies related to
gravitational waves from pulsars. 
The problem of detecting gravitational waves from known and unknown pulsars 
or rotating neutron stars was discussed. In the case of known pulsars 
 a general data analysis scheme for detecting 
gravitational waves from millisecond pulsars with resonant bar antennas was 
proposed \cite{san1}. As a specific example, the case of the nearest known
millisecond pulsar PSR~0437-4715 and the resonant  bar antenna at the
University of Western Australia was considered, since the gravitational wave 
frequency of the pulsar coincides with the resonant  frequency of the bar 
within a fraction of a Hertz.
The key idea was to rotate the phase plane with appropriate angular velocity
and thus correct for the  Doppler shift in the apparent pulsar frequency.  
For the University of Western Australia niobium antenna tuned to PSR 437 - 4715,  
astrophysically relevant sensitivity could be achieved with an improved 
transducer technology. It was concluded that the  best candidate for detection 
in this way is PSR 437 - 4715.  
 Also a list of pulsars was given, which  in principle could be
detected by resonant bar antennas operating between frequencies 500 Hz to
1300 Hz.

The other problem which is far more difficult is of detecting unknown pulsars 
or rotating neutron stars since in this case we do not know from electromagnetic 
observations the frequency or the location of the pulsar in the sky. 
Schutz \cite{bfs91}
has analysed the problem of the {\it all sky all frequency search} and shows 
that one may have to search for $10^{13}$ patches or directions in the sky 
for reasonable parameters. This could involve to the tune of $10^{23}$
floating point operations. Different schemes therefore need to be tried. 
One such  scheme using differential geometric techniques of scanning the pulsar
signal manifold with appropriate basis functions 
on the celestial sphere was proposed. 
A promising set of basis functions are the Gelfand functions. These functions 
have proved useful in the past in elegantly describing the antenna patterns of 
laser interferometric gravitational wave detectors and resonant bars 
\cite{dt88}.
\section{ Limit on rotation of relativistic stars}
{\sf John Friedman} summarised the state of our understanding of the
rotation limits of neutron stars.
The angular velocity of a uniformly rotating star could not exceed that of
a satellite in orbit at the star's equator.  This limiting {\it Kepler
frequency}, $\Omega_K$, was about half the frequency of a satellite at
the equator of the corresponding spherical star, because the star's
radius increased sharply as $\Omega$ approached $\Omega_K$.  For fixed
mass, its value was sensitive to the equation of state (EOS): a soft 
equation
of state implied a centrally condensed star with larger binding energy
and a correspondingly larger value of $\Omega_K$ than that of a stiff
EOS.  The uncertainty in the equation of state above nuclear density was
large, and the stiffest proposed EOSs gave stars with moments of
inertia at $\Omega=\Omega_K$ about 8 times greater for the stiffest
equations of state than for the softest consistent with our
knowledge. 

The limiting angular velocity could be reached only if the magnetic field
was small.  For neutron stars that were born with a sufficiently weak
magnetic field, a nonaxisymmetric instability probably set a limit
that was slightly more stringent limit than the Kepler frequency. 
The instability, however, was damped by bulk viscosity for temperatures 
above $2\times 10^{10}$ K and by an effective shear viscosity below 
the superfluid transition temperature.  As a result, it was not 
likely not to play a role in neutron stars spun up by accretion.

An observational upper limit on neutron star rotation could constrain the
equation of state of matter above nuclear density. In particular, the
two fastest of the known pulsars have angular velocities within $3\%$
 of
each other (about 640 Hz); if they turned out to be rotating at nearly
the upper limit on rotation for a 1.4 $M_\odot$ neutron star, then the 
equation of state above nuclear density was unexpectedly stiff.  

One could set an upper limit on spin that was independent of the (unknown)
equation of state above nuclear density.  Causality and the observation
of neutron stars (or, not to bias the case, against, say, quark stars
or stars with quark interiors) with mass at least as large as 1.44
$M_\odot$ constrained the equation of state.  With these minimal assumptions,
the minimum period of a relativistic star was 0.30 ms 
\cite{john1,john2,john3,john4}.

\section{Contributed Papers} 

\subsection{Theoretical} 

{\sf A. Gopakumar} \cite{GI96} 
 presented his work on the 2PN evolution 
of inspiralling compact binaries in general orbits using the post-Minkowskian 
approach \cite{BDI95}. 
They have computed the 2PN accurate mass quadrupole moment for a system of 
two point masses, and calculated the 2PN contributions to energy and angular 
momentum fluxes. Both the quadrupole moment and the energy loss agree with 
those obtained by Will and Wiseman using the Epstein-Wagoner approach. 
They have also calculated the 2PN waveforms for general orbits.  
The results are compared with 
those obtained using the perturbation methods .
 \subsection{Detectors and Technologies} 
 In this section the status of the prospects and the present and future 
sensitivity of both bar and laser interferometric detectors were discussed. 

 {\sf David Blair} 
from Australia reported on the work of his bar detector group.
He pointed out that in the high frequency regime $\approx 1000 Hz$ where the 
the resonant detectors operate they are now as sensitive as the LIGO-I and 
in two to three years they would achieve sensitivities a bit lower than 
the expected ones for LIGO-II (which would be at least a decade
 in the future). 
 
{\sf Massimo Bassan} 
from Rome  also pointed out the high sensitivity of the 
present generation of the bar detectors, which are the only operational 
detectors at present. An important point in his talk was the proposal for 
an array of 20-30 small detectors which would be sensitive in the bandwidth of 
a few kHz, since each one separately was not sensitive enough one needed a group 
of them for accumulating sensitivity \cite{FPB}. 
Nevertheless, such an array would use existing technology without the need of 
advanced cryogenics and the cost of such arrays would not be high. 
The arrays of detectors would 
be useful for the detection of burst signals from collapsing objects to 
neutron stars or small black holes. 
 
{\sf Ju Li} 
reported a promising result i.e. the use of sapphire for mirrors in 
the laser interferometers. Sapphire with its excellent mechanical and
thermal properties is an attractive material to use for beam splitter and
test masses in laser interferometer gravitational wave detectors.
The internal thermal noise amplitude of a sapphire test mass would be $16$
times better than that of a fused silica test mass with same dimensions.
If large sapphire samples maintained similar performance the total local
noise sources in laser interferometers could be reduced by factors of $16$. 
Consequently laser interferometric detectors with a few hundred meters arm
(AIGO400) would be comparable to that of the LIGO with arm length of 4 Kms
or alternatively the kilometre armlength interferometers would have
enhanced sensitivity. The uncertainity principle sets the fundamental
limit which is almost close to the limits envisaged 
\cite{blair1,blair2,blair3}.
 
{\sf Vijay Chickarmane} 
discussed the possibility of enhancing the sensitivity of a dual recycled 
interferometer (with Fabry-Perot cavities in the arms), using the squeezed 
state technique \cite{CD96}. 
They  calculated the sensitivity for both the broad and narrow band modes 
of operation of the interferometer. 
The result was that  in the broad band case the squeezing was not so useful,
however, in the narrow band case squeezed light could be used to enhance the 
sensitivity. For that case of 60\% squeezing, the gain in sensitivity was by a 
factor 2.5. 
 
{\sf David McClelland} 
also discussed  the dual recycling and  reported on `bench top' recycling 
experiments underway at the Australian National University. 
He pointed out the importance of the dual recycling for mid-baseline
gravitational wave detectors 
like GEO600, TAMA300 and AIGO400 since with the combination of innovative
suspension systems and optical materials they would be able to achieve 
sensitivities in excess of first stage  LIGO and VIRGO over a bandwidth of 
a few hundred Hertz. 
 
{\sf Gabriella Gonzalez} 
discussed the motivation and current status of
the MIT phase noise experiment. 
 Above $200$ Hz the sensitivity
is determined by how well the phase difference between the two beams can
be determined at the detector. Given the limit of photon shot noise 
LIGO would require $70$ Watts of laser light incident on the beam splitter to
reach the required sensitivity. To study noise at high power levels a team
at MIT was putting together a recycled Michelson interferometer. 
The system was a asymmetric single bounce Michelson
interferometer with recycling gain $100$ incorporating on a small scale the 
LIGO suspension design and active and passive LIGO seismic isolation.

\subsection{Data Analysis} 

{\sf R. Balasubramanian}  
discussed  strategies for analyzing 
data from coalescing binaries and presented  results of Monte Carlo 
estimation of the parameters of the binary system \cite{bsd95,bsd96}.
He presented a formalism using differential geometric techniques
which generalises the 
problem of choosing an optimal set of filters to detect the chirp waveform. 
He   pointed out the need for finding  sets of convenient parameters and 
he  showed that even after the inclusion of 2PN corrections, the waveform 
could essentially be detected by using a one-dimensional lattice of templates. 
He also presented  the results of a Monte Carlo simulation using the above 
formalism. The results of the simulations have shown that the covariance 
matrix underestimates the actual errors in the estimation of parameters, even 
when the signal to noise ratio is as high as 20. He concluded that since the 
detection of events with high signal to noise ratio  would be very rare the 
covariance matrix was inadequate to describe the errors in the measurement of 
the parameters of the waveform. 
 
{\sf Piotr Jaranowski} 
presented his work \cite{JK94,JKKT96} on the estimation of parameters of 
the gravitational-wave signal from a coalescing binary by a network of laser
interferometers. The solution of the inverse problem (using maximum 
likelihood and least squares methods) for the network of 3 detectors was
generalized to the network of $N$ detectors. This enabled, from measurements 
at individual detectors of the network, optimal estimation of the 
astrophysically interesting parameters of the binary system:  its distance 
from Earth, its position in the sky, and the chirp mass of the system. 
The accuracy of the estimation of the parameters was assessed from the 
inverse
of the Fisher information matrix. Extensive Monte Carlo simulations were 
performed to assess the accuracy of the estimation of the astrophysical 
parameters by networks of 3 and 4 detectors. He  reported that the addition 
of the fourth node to LIGO/VIRGO network in Australia  increases the number 
of detectable events roughly by two times and accuracy of position 
determination by 3-4 times.
 
{\sf S.D. Mohanty}  
presented an alternative method for analysing data 
from coalescing binaries, which is a {\em modified periodogram} \cite{MS96}
and discussed the scaling laws for chirps.
In this approach all the features of the standard matched filtering appear 
nevertheless it was a less time consuming method which had the advantage of 
being statistically independent for the same sample of noisy data. 
 
{\sf Kanti Jotania}  
talked about the analysis of gravitational waves from 
pulsars \cite{JVD96}. 
He mentioned  various obstacles related to this detection since
the signals are expected to be very weak and thus the observation times 
should be long (a few months).
This created additional problems since a monochromatic signal
became both frequency and amplitude 
modulated due to the rotation and the orbital 
motion of the earth.
The effect of these two modulations was to smear out the 
monochromatic signal into a small bandwidth about the signal frequency of the 
wave.
He  showed the results of his study on the Fourier transform of the 
pulsar signal  taking into account the rotation of the Earth for one day 
observational period. 
Finally, he   showed an analytic form of the Fourier transform 
considering the rotation of the Earth including the
orbital corrections. 
 
{\sf Biplab Bhawal}  
discussed the coincidence detection of broadband 
signals by the planned interferometric gravitational wave detectors 
\cite{BD96}. 
He had taken into account the six planned detectors (2 LIGOs, VIRGO, GEO600, 
AIGO400, TAMA300) and performed coincidence experiments for the detection of 
broadband signals coming either from coalescing compact binaries or burst 
sources. He  showed results on the comparisons of the achievable 
sensitivities of these detectors under different optimal configurations and 
found that a meaningful coincidence experiment could only be performed by a 
network where the LIGOs and VIRGO are operated in power recycling mode and 
other medium scale detectors (GEO, AIGO, TAMA) are operated in dual recycling 
mode with a narrower bandwidth. Finally, for effectively optimizing the values 
for different possible networks he 
calculated the time-delay window sizes. The effect of filtering on
calculation of thresholds and the volume of sky covered by the networks
were also obtained.

\end{document}